%% file: main.tex
\documentclass[sigconf]{aamas}  
\pdfoutput=1
\usepackage{booktabs}

\usepackage{caption}
\usepackage{siunitx}

\begin{document}

\title{Learning Reciprocity in Complex Sequential Social Dilemmas}  





\author{Tom Eccles}
\affiliation{%
 \institution{DeepMind}
 \city{London} 
 \country{UK} 
}
\email{eccles@google.com}
\author{Edward Hughes}
\affiliation{%
 \institution{DeepMind}
 \city{London} 
 \country{UK} 
}
\email{edwardhughes@google.com}
\author{J\'{a}nos Kram\'{a}r}
\affiliation{%
 \institution{DeepMind}
 \city{London} 
 \country{UK} 
}
\email{janosk@google.com}
\author{Steven Wheelwright}
\affiliation{%
 \institution{DeepMind}
 \city{London} 
 \country{UK} 
}
\email{sjwheel@google.com}
\author{Joel Z. Leibo}
\affiliation{%
 \institution{DeepMind}
 \city{London} 
 \country{UK} 
}
\email{jzl@google.com}

%
%
%
%
%
%

\begin{abstract}  
Reciprocity is an important feature of human social interaction and underpins our cooperative nature. What is more, simple forms of reciprocity have proved remarkably resilient in matrix game social dilemmas. Most famously, the tit-for-tat strategy performs very well in tournaments of Prisoner's Dilemma. Unfortunately this strategy is not readily applicable to the real world, in which options to cooperate or defect are temporally and spatially extended. Here, we present a general online reinforcement learning algorithm that displays reciprocal behavior towards its co-players. We show that it can induce pro-social outcomes for the wider group when learning alongside selfish agents, both in a $2$-player Markov game, and in $5$-player intertemporal social dilemmas. We analyse the resulting policies to show that the reciprocating agents are strongly influenced by their co-players' behavior.
\end{abstract}

%


\maketitle


\input{body.tex}


\bibliographystyle{ACM-Reference-Format}  
\bibliography{sample-bibliography.bib}  

\input{appendix.tex}
\end{document}

%% file: body.tex
\section{Introduction}

Sustained cooperation among multiple individuals is a hallmark of human social behavior, and may even underpin the evolution of our intelligence \cite{Reader4436, dunbar_1993}. Often, individuals must sacrifice some personal benefit for the long-term good of the group, for example to manage a common fishery or provide clean air. Logically, it seems that such problems are insoluble without the imposition of some extrinsic incentive structure \cite{olson1965logic}. Nevertheless, small-scale societies show a remarkable aptitude for self-organization to resolve public goods and common pool resource dilemmas \cite{ostrom1990governing}. Reciprocity provides a key mechanism for the emergence of collective action, since it rewards for pro-social behavior and punishes for anti-social acts. Indeed, it is a common norm shared by diverse societies \cite{becker1990reciprocity, ostrom_1998, blau1964exchange, thibaut1966social}. Moreover, laboratory studies find experimental evidence for conditional cooperation in public goods games; see for example \cite{CROSON200595}. 

By far the most well-known model of reciprocity is Rapoport's Tit-for-Tat \cite{chammah1965prisoner}. This hand-coded algorithm was designed to compete in tournaments where each round consisted of playing the repeated Prisoner's Dilemma game against an unknown opponent. The algorithm cooperates on its first move, and thereafter mimics the previous move of its partner, by definition displaying perfect reciprocity. Despite its simplicity, Tit-for-Tat was victorious in the tournaments \cite{10.2307/173932, 10.2307/173638}. Axelrod \cite{robert1984evolution} identified four key features of the Tit-for-Tat strategy which contributed to its success, which later successful algorithms such as win-stay lose-shift \cite{Nowak1993} also share:
\begin{itemize}
    \item Nice: start by cooperating.
    \item Clear: be easy to understand and adapt to.
    \item Provocable: retaliate against anti-social behavior.
    \item Forgiving: cooperate when faced with pro-social play.
\end{itemize}
Later, we will think of these as design principles for models of reciprocity.

Although Tit-for-Tat and its variants have proved resilient to modifications in the matrix game setup \cite{DBLP:journals/corr/abs-1301-5683, BOYD198947, Nowak2006}, it is clearly not applicable to realistic situations. In general, cooperating and defecting require an agent to carry out complex sequences of actions across time and space, and the payoffs defining the social dilemma may be delayed. More sophisticated models of reciprocity should be applicable to the multi-agent reinforcement learning domain, where the tasks are defined as Markov games \cite{Shapley1095, Littman:1994:MGF:3091574.3091594}, and the nature of cooperation is not pre-specified. In this setting, agents must learn both the high-level strategy of reciprocity and the low level policies required for implementing (gradations of) cooperative behavior. An important class of such games are intertemporal social dilemmas \cite{DBLP:journals/corr/PerolatLZBTG17, DBLP:journals/corr/abs-1803-08884}, in which both individual rationality is at odds with group-level outcomes and the negative impact of individual greed is temporally distant from its adverse consequences for the group.

\cite{DBLP:journals/corr/LererP17, kleiman, DBLP:journals/corr/abs-1710-06975} propose reinforcement learning models for $2$-agent problems, based on a planning approach. Both models first pre-train cooperating and defecting policies using explicit knowledge of other agents' rewards. The policies are then used as options in hand-coded meta-policies. The main variation between these approaches is the algorithm for switching between these options. In \cite{DBLP:journals/corr/LererP17} and \cite{kleiman}, the decision about which policy is chosen in response to the last action, and that action is assessed through planning. In \cite{DBLP:journals/corr/abs-1710-06975}, the decision is based on the recent rewards of the agent, which defects if it is not doing sufficiently well. These models of reciprocity are important stepping stones, but have some important limitations. Firstly, reciprocity imitates a range of behaviors, rather than $2$ pre-determined ones (pure cooperation and pure defection), which are not obviously well-defined in general. Secondly, reciprocity is applicable beyond the $2$-player case. Thirdly, reciprocity does not necessarily rely on directly observing the rewards of others. Finally, reciprocity can be learned online, offering better scalability and flexibility than planning.

We propose an online-learning model of reciprocity which addresses these limitations while still significantly outperforming selfish baselines in the $2$-player setting. Our setup comprises two types of reinforcement learning agents, \textit{innovators} and \textit{imitators}. Both innovators and imitators are trained using the A3C algorithm \cite{DBLP:journals/corr/MnihBMGLHSK16} with the VTrace correction \cite{DBLP:journals/corr/abs-1802-01561}. An innovator optimizes for a purely selfish reward. An imitator has two components: (1) a mechanism for measuring the level of sociality of different behaviors and (2) an intrinsic motivation \cite{NIPS2004_2552} for matching the sociality of others. We investigate two mechanisms for assessing sociality. The first is based on hand-crafted features of the environment. The other uses a learned ``niceness network'', which estimates the effect of one agent's actions on another agent's future returns, hence providing a measure of social impact \cite{latane}. The niceness network also encodes a social norm among the imitators, for it represents ``a standard of behavior shared by members of a social group'' \cite{norm_enc}. Hence our work represents a model-based generalization of \cite{Sen:2007:ENT:1625275.1625519} to Markov games.

An innovator's learning is affected by the reciprocity displayed by imitators co-training in the same environment. The innovator is incentivized to behave pro-socially, because otherwise its anti-social actions are quickly repeated by the group, leading to a bad payoff for all, including the innovator. With one innovator and one imitator, the imitator learns to respond in a Tit-for-Tat like fashion, which we verify in a dyadic game called the \textit{Coins} dilemma \cite{DBLP:journals/corr/abs-1710-06975}. For one innovator with many imitators, the setup resembles the phenomenon of pro-social leadership \cite{Henrich20150013, GACHTER2018321}. Natural selection favours altruism when individuals exert influence over their followers; we see an analogous effect in the reinforcement learning setting. 

More specifically, we find the presence of imitators elicits socially beneficial outcomes for the group (5 players) in both the \textit{Harvest} (common pool resource appropriation) and \textit{Cleanup}  (public good provision) environments \cite{DBLP:journals/corr/abs-1803-08884}. We also quantify the social influence of the innovator on the imitators by training a graph network \cite{battaglia2016interaction} to predict future actions for all agents, and examining the edge norms between the agents, just as in \cite{tacchetti2018relational}. This demonstrates that influence of the innovator on the imitator is greater than the influence between other pairs of agents in the environment. Moreover, we find that the innovator's policies return to selfish free-riding when we continue training without reciprocating imitators, showing that reciprocity is important for maintaining stability of a learning equilibrium. Finally, we demonstrate that the niceness network learns an appropriate notion of sociality in the dyadic \textit{Coins} environment, thereby inducing a tit-for-tat like strategy.

\section{Agent models}

We use two types of reinforcement learning agent, \emph{innovators} and \emph{imitators}. Innovators learn purely from the environment reward. Imitators learn to match the sociality level of innovators, hence demonstrating reciprocity. We based the design of the imitators on Axelrod's principles, which in our language become:
\begin{itemize}
    \item Nice: imitators should not behave anti-socially at the start of training, else innovators may not discover pro-sociality.

    \item Clear: imitation must occur within the timescale of an episode, else innovators will be unable to adapt.
    
    \item Provocable: imitators must reciprocate anti-social behavior from innovators.
    
    \item Forgiving: the discount factor with which anti-social behavior is remembered must be less than $1$.
\end{itemize}

Note that imitating the policy of another agent over many episodes is \emph{not} sufficient to produce cooperation. This type of imitation does not change behaviour during an episode based on the other agent's actions, and so does not provide feedback which enables the innovators to learn. We validate this in an ablation study. As such our methods are complementary to, but distinct from the extensive literature on imitation learning; see \cite{Hussein:2017:ILS:3071073.3054912} for a survey. 

Moreover, observe that merely training with collective reward for all agents is inferior to applying reciprocity in several respects. Firstly collective reward suffers from a lazy agent problem due to spurious reward \cite{Sunehag:2018:VNC:3237383.3238080}. Secondly, it produces agent societies that are exploitable by selfish learning agents, who will free-ride on the contributions of others. Finally, in many real-world situations, agents do not have direct access to the reward functions of others. 

\subsection{Innovator}
The innovator comprises a deep neural network trained to generate a policy from observations using the asynchronous advantage actor-critic algorithm \cite{DBLP:journals/corr/MnihBMGLHSK16} with V-Trace correction for stale experience \cite{DBLP:journals/corr/abs-1802-01561}. For details of the architecture, see Appendix \ref{appendix:arch}.

\subsection{Imitator}

We use two variants of the imitator. The difference is in what is being imitated. The \emph{metric-matching imitator} imitates a hand-coded metric. The \emph{niceness network imitator} instead learns what to imitate. The metric-matching variant allows for more experimenter control over the nature of reciprocity, but at the expense of generality. Moreover, this allows us to disentangle the learning dynamics which result from reciprocity from the learning of reciprocity itself, a scientifically useful tool. On the other hand, the niceness network can readily be applied to any multi-agent reinforcement learning environment with no prior knowledge required.

\subsubsection{Reinforcement learning}
Imitators learn their policy using the same algorithm and architecture as innovators, with an additional intrinsic reward to encourage reciprocation. Consider first the case with $1$ innovator and $1$ imitator; the general case will follow easily. Let the imitator have trajectory $T_{\textrm{im}}$, and the innovator has trajectory $T_{\textrm{inv}}$. Then the intrinsic reward is defined as
\begin{equation}\label{eq:intrinsic_reward}
    r_{\textrm{int}}(t) = - (N(T_{\textrm{im}}) - N(T_{\textrm{inv}}))^2 \, ,
\end{equation}
where $N(T)$ is some function of the trajectory, which is intended to capture the effect of the actions in the trajectory on the return of the other agent. We shall refer to $N(T)$ as the \textit{niceness} of the agent whose trajectory is under consideration. This intrinsic reward is added to the environment reward. We normalize the intrinsic reward so that it accounts for a proportion $c_{\textrm{im}}$ of the total absolute reward in each batch:
\begin{equation}
    r(t) = r_{\textrm{env}}(t) + c_{\textrm{im}}\, \frac{r_{\textrm{int}}(t)}{\mu(r_{\textrm{int}})} \, ,
\end{equation}
where the mean is taken over a batch of experience and $c_{im}$ is a constant hyperparameter, which we tuned separately on each environment.

Generalizing to the case of $1$ innovator with $M$ imitators is simple: we merely apply the intrinsic reward to each imitator separately, based on the deviation between their niceness and that of the innovator. Since our method uses online learning, it automatically adapts to the details of multi-agent interaction in different environments. This is difficult to capture in planning algorithms, because the correct cooperative policy for interactions with one agent depends on the policies of all the other agents.

The two imitator variants differ primarily in the choice of the niceness function $N(\cdot)$, as follows.

\subsubsection{Metric matching}
For the metric-matching imitator, we hand-code $N(T)$ for trajectories in each environment in a way that measures the pro-sociality of an agent's behavior. If these metrics are accurate, this should lead to robust reciprocity, which gives us a useful comparison for the niceness network imitators.

\subsubsection{Niceness network}
The niceness network estimates the value of the innovator's states and actions to the imitator. Let $V(t)$ be the discounted return to the imitator from time $t$. We learn approximations to the following functions:
\begin{align*}
    V^{\pi_{\textrm{inv}}}(s^{\textrm{inv}}_t) &= \mathbb E(V(t) | s^{\textrm{inv}}_t) \, ,\\
    Q^{\pi_{\textrm{inv}}}(s^\textrm{inv}_t, a^{\textrm{inv}}_t) &= \mathbb E(V(t) | s^{\textrm{inv}}_t, a^{\textrm{inv}}_t) \, ,
\end{align*}
where $s^{\textrm{inv}}_t$ and $a^{\textrm{inv}}_t$ are the state and action of the innovator at time $t$. Clearly this requires access to the states and actions of the innovator. This is not an unreasonable assumption when compared with human cognition; indeed, there is neuroscientific evidence that the human brain automatically infers the states and actions of others \cite{Mitchell1309}. Extending our imitators to model the states and actions of innovators would be an valuable extension, but is beyond our scope here.

The niceness of action $a^{\textrm{inv}}_t$ is defined as
\begin{equation}
    n^{\pi_{\textrm{inv}}}(s^{\textrm{inv}}_t, a^{\textrm{inv}}_t) = Q^{\pi_{\textrm{inv}}}(s^{\textrm{inv}}_t, a^{\textrm{inv}}_t) - V^{\pi_{\textrm{inv}}}(s^{\textrm{inv}}_t) \, .
\end{equation}
This quantity estimates the expected increase in the discounted return of the imitator given the innovator's action $a^{\textrm{inv}}_t$.

Then, for a generic trajectory $T=(s_1, a_1, \dots s_t, a_t)$ we define the niceness of the trajectory, $N^{\pi_{\textrm{inv}}}(T)$, to be
\begin{equation}
    N^{\pi_{\textrm{inv}}}(T) = \sum_{i=1}^{t} \gamma^{t-i} n^{\pi_{\textrm{inv}}}(s_i, a_i).
\end{equation}
This $N^{\pi_{\textrm{inv}}}(T)$ is used in as the function $N$ in calculating the intrinsic reward \ref{eq:intrinsic_reward}

The parameter $\gamma$ controls the timescale on which the imitation will happen; the larger $\gamma$ is, the slower the imitator is to forget. This balances between the criteria of provocability and forgiveness.

We learn the functions $V^{\pi_{\textrm{inv}}}$ and $Q^{\pi_{\textrm{inv}}}$ using the $\textrm{TD}(\lambda)$ algorithm \cite{Sutton:1998:IRL:551283} using the innovator's states and actions and the imitator's reward.

While the niceness network is trained only on the states and actions of the innovator, in calculating the intrinsic reward it is used for inference on both imitator and innovator trajectories. For this to work we require symmetry between the imitator and the innovator: they must have the same state-action space, the same observation for a given state and the same payoff structure. Therefore our approach is not applicable in asymmetric games. Cooperation in asymmetric games may be better supported by indirect reciprocity than by generalizations of Tit-for-Tat; see for example \cite{pmid17956841}.

\subsubsection{Off-policy correction}
When calculating the intrinsic reward for the niceness network imitator, we evaluate $N^{\pi_{\textrm{inv}}}(T_{\textrm{im}})$ for the imitator's trajectories, having only trained on trajectories from the innovator. This is problematic if the states and actions of $T_{\textrm{im}}$ are drawn from a different distribution from those of $T_{\textrm{inv}}$. In this case, on the trajectory $T_{\textrm{im}}$, we might expect that $V^{\pi_{\textrm{inv}}}$ and $Q^{\pi_{\textrm{inv}}}$ would be inaccurate estimates for the effect of the imitator on the innovator's payoff. In other words, the flip of perspective from innovator to imitator at inference time is only valid if the imitator's policy is sufficiently closely aligned with that of the innovator.

In practice, we find that applying $Q^{\pi_{\textrm{inv}}}$ to $T_{\textrm{im}}$ is particularly problematic. Specifically if the imitator's policy contains actions which are very rare in the innovator's policy, then the estimate of $Q^{\pi_{\textrm{inv}}}$ for these actions is not informative. To correct this issue, we add a bias for policy imitation to the model. The imitator estimates the policy of the innovator, giving an estimate $\hat\pi_{\textrm{inv}}(s)$ for each state $s$. Then we add an additional KL-divergence loss for policy imitation,
\begin{equation}
    L = c_{\textrm{KL}}D_{\textrm{KL}}(\pi_{\textrm{im}}(s), \hat\pi_{\textrm{inv}}(s)) \, ,
\end{equation}
where $\pi_{\textrm{im}}(s)$ is the policy of the imitator, and $c_{\textrm{KL}}$ is a constant hyperparameter. The effect of this loss term is to penalize actions that are very unlikely in $\pi_{\textrm{inv}}(s)$; these are the actions for which the niceness network is unable to produce a useful estimate.

In our ablation study, we will demonstrate that the policy imitation term alone does not suffice to yield cooperation. Indeed, our choice of terminology was based on the high-level bias for reciprocity introduced by the niceness matching, not the low-level policy imitation. The latter might better be thought of as an analogue to human behavioral mimicry \cite{chartrand2013antecedents}, which while an important component of human social interaction \cite{lakin2003chameleon}, alone does not constitute a robust means of generating positive social outcomes \cite{HALE2016106}.

Empirically we find that niceness matching alone often suffices to generate cooperative outcomes, despite the off-policy problem. This is likely because environments contain some state-action pairs which are universally advantageous or disadvantageous to others, regardless of the particular policy details. Moreover, this is not a limitation of our environments, it is a feature familiar from everyday life: driving a car without a catalytic converter is bad for society, regardless of the time and place. The policy imitation correction does however serve to stabilize cooperative outcomes, a suggested effect of mimicry in human social groups \cite{Tanner2008OfCA}.

\section{Experiments}

We test our imitation algorithms in three domains. The first is Coins. This is a $2$-player environment introduced in \cite{DBLP:journals/corr/LererP17}. This environment has simple mechanics, and a strong social dilemma between the two players, similar to the Prisoner's Dilemma. This allows us to study our algorithms in a setup close to the Prisoner's Dilemma, and make comparisons to previous work.
    
The other two environments are Harvest and Cleanup. These are more complex environments, with delayed results of actions, partial observability of a somewhat complex gridworld, and more than two players. These environments are designed to test the main hypothesis of this paper, that our algorithms are able to learn to reciprocate in complex environments where reciprocity is temporally extended and hard to define. We choose these two environments because they represent different classes of social dilemma; Cleanup is a public goods game, while Harvest is a common pool resource dilemma.

\subsection{Coins}
\subsubsection{Environment}

\begin{figure}
 \centering
 \includegraphics[scale=0.7]{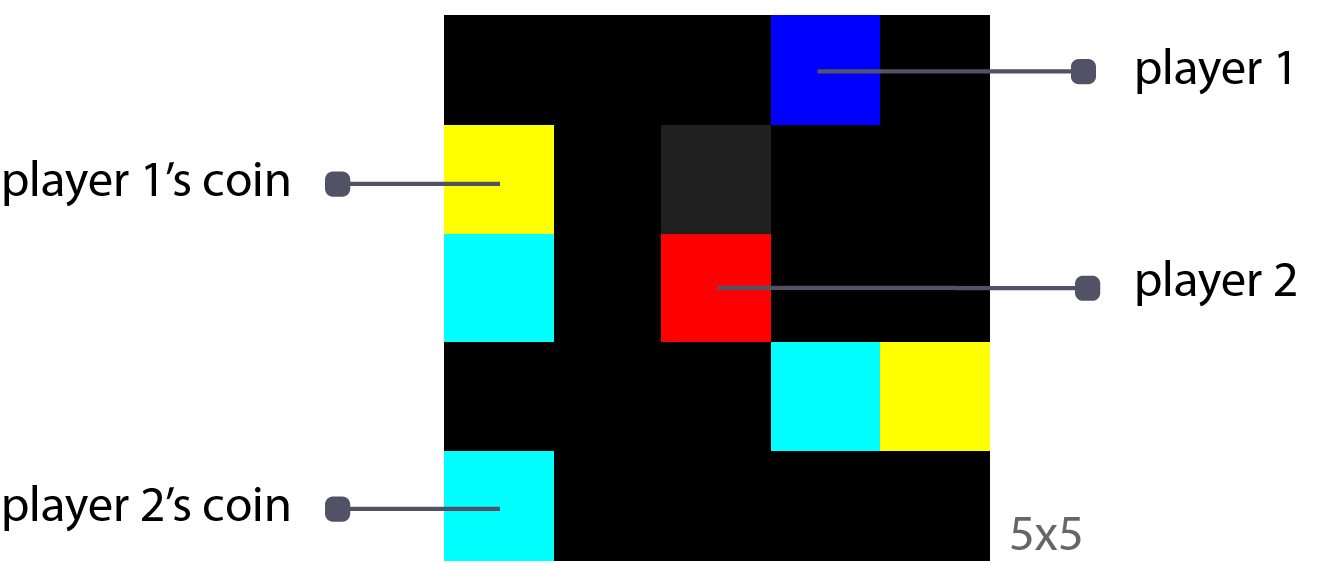}
\caption{The \textit{Coins} game. Agents are rewarded for picking up coins, and punished when the other agent picks up a coin of their color.}
\end{figure}

We use the gridworld game Coins, introduced in \cite{DBLP:journals/corr/LererP17}. Here, two players move on a fully-observed $5\times5$ gridworld, on which coins of two colors periodically appear. When a player picks up a coin of either color, they get a reward of $1$. When a player picks up a coin of the other player's color, the other player gets a reward of $-2$. The episode ends after $500$ steps. The total return is maximized when each player picks up only coins of their own color, but players are tempted to pick up coins of the wrong color. At each timestep coins spawn each unoccupied square with a probability $0.005$. Therefore the maximum achievable collective return is approximately $50$ in expectation, if neither agent chooses to defect and both agents collect all coins of their own color.

In this game, the metric-matching agent uses the number of recent defections as its measure of niceness $N(T)$. We define $n(s_t, a_t)$ to be $-1$ if the action $a_t$ picks up a coin which penalizes the other player, and $0$ otherwise. Then we define
\begin{equation}
    N(T) = \sum_{i=1}^{t} \lambda^{t-i} n(s_{i}, a_{i}).
\end{equation}
To make the environment tractable for our niceness network we symmetrize the game by swapping the coin colors in the observation of the innovator. This means that the value and $Q$ functions for the innovator can be consistently applied for the imitator. Note that only the colors in the observation are swapped; the underlying game dynamics remain the same, so the social dilemma still applies.

\subsubsection{Results}
\begin{figure}
 \centering
 \includegraphics[scale=0.4]{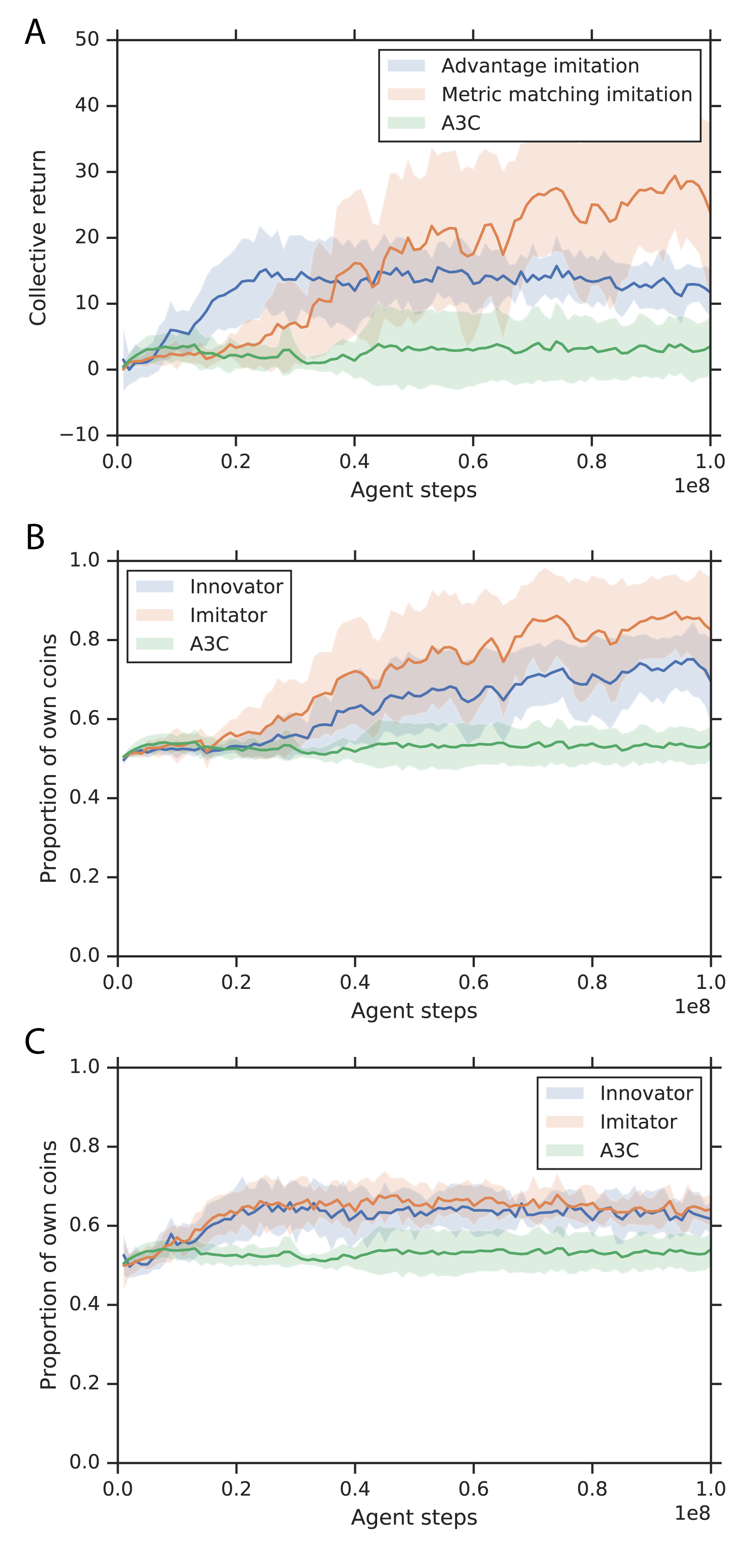}
\caption{Reciprocity generates pro-social outcomes in Coins. (A) Both metric-matching and niceness-network variants significantly outperform the baseline, measured according to the total payoff. (B) The metric-matching imitators closely match the coin collection profile of the innovator during training, driving the innovator towards pro-sociality. No such beneficial outcome is seen for the selfish agents. (C) The same holds for the niceness network imitators.}
\label{fig:coins_results}
\end{figure}

Both the metric-matching and niceness network imitators outperformed the greedy baseline in this environment, reaching greater overall rewards and a larger proportion of coins being collected by the agent which owns them (Figure \ref{fig:coins_results}). However, neither model was able to reach the near-perfect cooperation achieved by the approximate Markov Tit-for-Tat algorithm \cite{DBLP:journals/corr/LererP17}, as shown in Figure 3 of that paper. We do not provide a numerical comparison, because we reimplemented the Coins environment for this paper.

This suggests that when it is possible to pre-learn policies that purely cooperate or defect and roll these out into the future, it is advantageous to leverage this prior knowledge to generate precise and extreme reciprocity. One might imagine improving our algorithm to display binary imitation behavior by attempting to match the maximum and minimum of recent niceness, rather than a discounted history. It would be interesting to see whether this variant elicited more cooperative behavior from innovators.

\subsection{Cleanup}
\subsubsection{Environment}
In the Cleanup environment, the aim is to collect apples. Each apple collected provides a reward of $1$ to the agent which collects it. Apples spawn at a rate determined by the state of a geographically separated river. Over time, this river fills with waste, lowering the rate of apple spawning linearly. For high enough levels of waste, no apples can spawn. The episode starts with the level of waste slightly above this critical point. The agents can take actions to clean the waste when near the river, which provides no reward but is necessary to generate any apples. The episode ends after $1000$ steps, and the map is reset to its initial state. For details of the environment hyperparameters, and evidence that this is a social dilemma, see \cite{DBLP:journals/corr/abs-1803-08884}.

More precisely, this is a public goods dilemma. If some agents are contributing to the public good by clearing waste from the river, there is an incentive to stay in the apple spawning region to collect apples as they spawn. However, if all players adopt this strategy, then no apples spawn and there is no reward.

\begin{figure}
 \centering
 \includegraphics[scale=0.4]{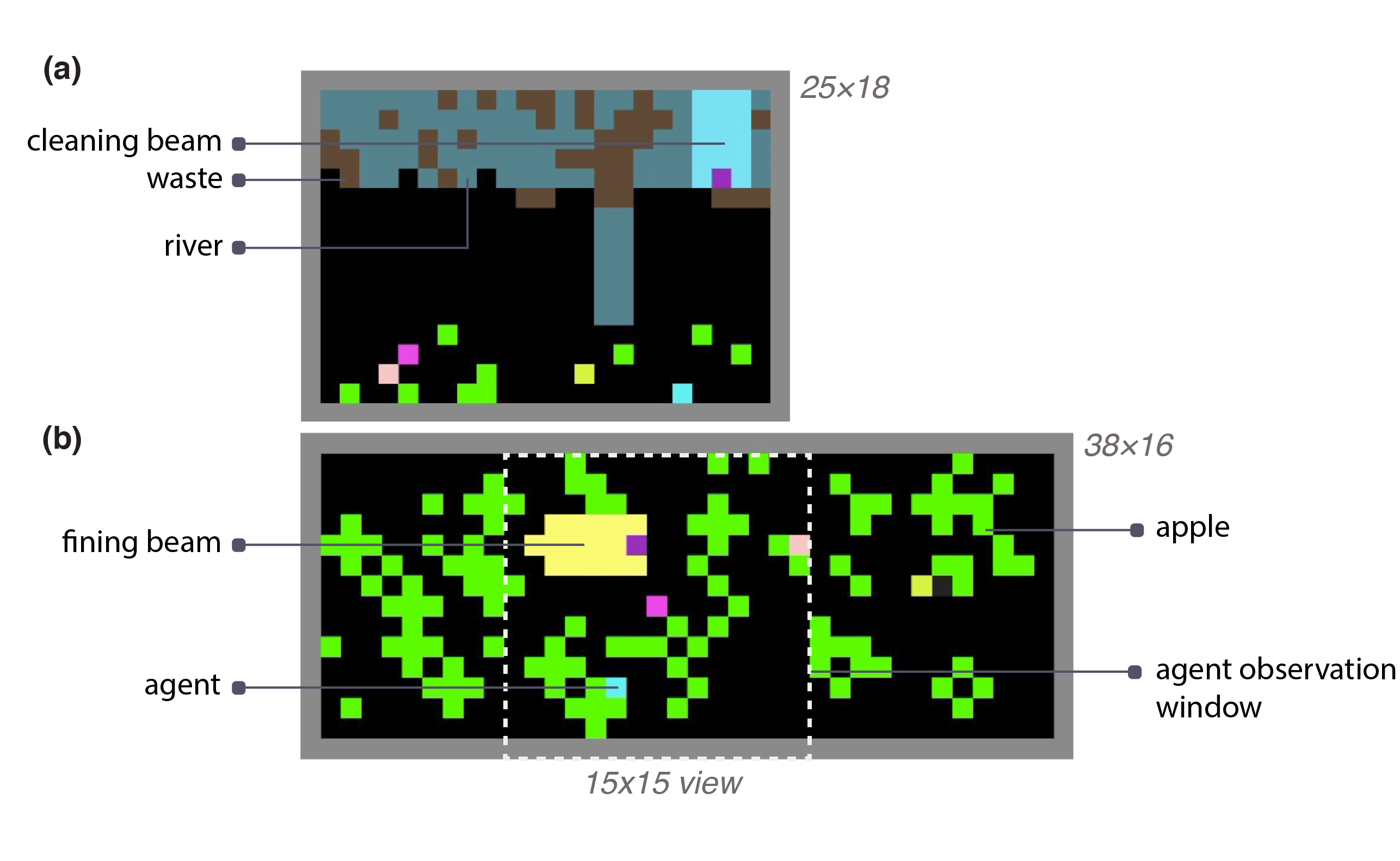}
\caption{The \textit{Cleanup} and \textit{Harvest} games. Agents take simultaneous actions in a partially observable $2$D gridworld. Rewards are obtained by eating apples, and agents may fine each other, conferring negative utility. The games are intertemporal social dilemmas: short term individual rationality is at odds with the long-term benefit of the whole group.}
\end{figure}

In this game, the metric-matching agent uses the number of contributions to the public good as its measure of niceness---for a given state and action, $n(s_t, a_t)$ is $1$ if the action $a_t$ removes waste from the river, and $0$ otherwise. Then we define
\begin{equation}
    N(T) = \sum_{i=1}^{t} \gamma^{t-i} n(s_{i}, a_{i}).
\end{equation}

\subsubsection{Results}
\begin{figure}
 \centering
 \includegraphics[scale=0.4]{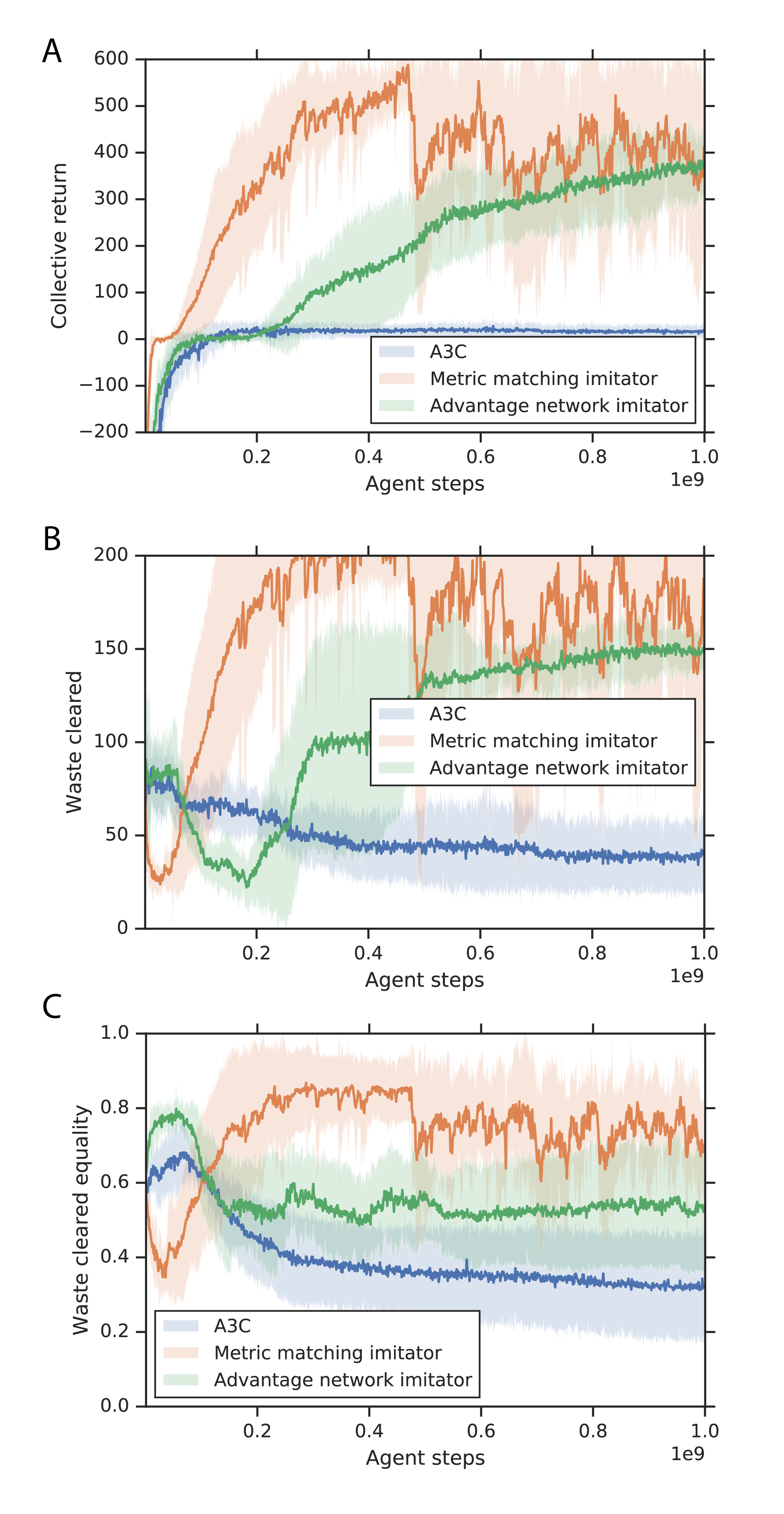}
\caption{The effect of reciprocity on social outcomes in Cleanup. (A) Collective return is higher when metric-matching or niceness-network imitators co-learn with a selfish innovator. (B) The improvement in collective return is mediated by a greater contribution to the public good. (C) Contributions remain reasonably equal in the imitation conditions, indicating that the imitators are successfully matching their pro-social behavior with that of the innovator.}
\label{fig:huangpu-results}
\end{figure}

Both metric-matching and niceness network imitators are able to induce pro-social behaviour in the innovator they play alongside, greatly exceeding the return and contributions to the public good of selfish agents (Figure \ref{fig:huangpu-results}). Niceness network imitators come close to the final performance of metric-matching imitators, despite having to learn online which actions are pro-social. A representative episode from the niceness network case reveals the mechanism by which the society solves the social dilemma.\footnote{A video is available at \url{https://youtu.be/kCjYfdVlLC8}.} The innovator periodically leads an expedition to clean waste, which is subsequently joined by multiple imitators. Everyone benefits from this regular cleaning, since many apples are made available (and consumed) throughout the episode.

\subsection{Harvest}
\subsubsection{Environment}
In the Harvest environment, introduced in \cite{DBLP:journals/corr/PerolatLZBTG17}, collecting apples again provides a reward of $1$. Harvested apples regrow at a rate determined by the number of nearby apples---the more other apples are present, the more like the apple is to regrow on a timestep. If all the apples in a block are harvested, none of them will ever regrow. The episode ends after $1000$ steps, and the map is reset to its initial state. For details of the environment hyperparameters, and evidence that this is a social dilemma, see \cite{DBLP:journals/corr/abs-1803-08884}.

This is a commons dilemma \cite{Hardin1243}. A selfish agent will harvest as rapidly as possible; if the whole group adopts this approach, then the resource is quickly depleted and the return over the episode is low. In order to get a high group return, agents must abstain from harvesting apples which would overexploit the common pool resource.

In this game, there is no clear metric of pro-sociality for the metric-matching agent we use. The anti-social actions in this game are taking apples with few neighbours, as this leads to the common pool resource being depleted. The sustainability of taking a particular apple $a$ can therefore be approximated by the number of apples within $\ell_1$ distance of $2$ to $a$, capped at $3$. We call this quantity $\textrm{sus}(a)$, following an analogous definition in \cite{janssen}. For a trajectory $T$ where an agent eats apples $(a_1, \dots, a_k)$ in order, we define
\begin{equation}
    N(T) = \sum_{i=1}^{k} \gamma^{k-i} \textrm{sus}(a_i) \, .
\end{equation}

\subsubsection{Results}
\begin{figure}
 \centering
 \includegraphics[scale=0.4]{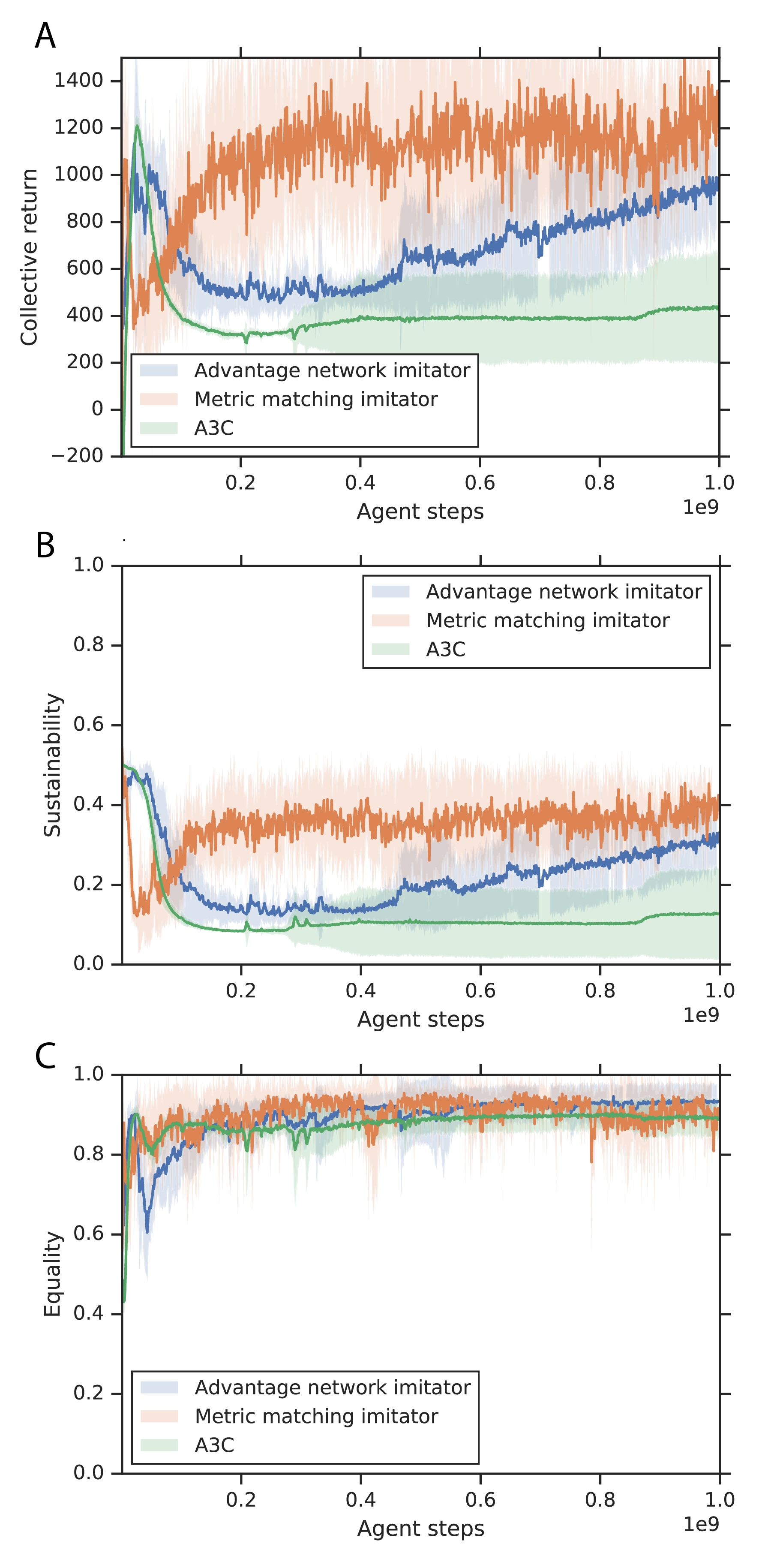}
\caption{The effect of reciprocity on social outcomes in Harvest.\protect\footnotemark\ (A) Collective return is higher when metric-matching or niceness-network imitators co-learn with a selfish innovator. (B) In the imitation conditions, the group learns a more sustainable strategy. (C) Equality remains high throughout training, suggesting that the imitators are successfully matching the cooperativeness of innovators.}
\label{fig:tragedy-results}
\end{figure}

We present our findings in Figure \ref{fig:tragedy-results}. Selfish agents are successful near the start of training, but as they become more efficient at collection they deplete the common pool resource earlier in each episode, and collective return falls. This effect is most obvious when examining the sustainability metric, which we define as the average proportion of the episode that had passed when each apple was collected. Agents which collect apples perfectly uniformly would achieve a sustainability of $0.5$. The baseline achieves a mere $0.1$.

For niceness network imitators, we see the same pattern near the start, where all the agents become more competent at collecting apples and begin to deplete the apple pool faster. We then see sustainability and collective return rise again. This is because the niceness network learns to classify sustainable behaviour, generating imitators that learn to match the sustainability of innovators, which creates an incentive for innovators to behave less selfishly. 

Similarly, the experiment with metric-matching imitators enters a tragedy of the commons early in training, before recovering to achieve higher collective return and better sustainability than the niceness network in a shorter period of training time. This makes intuitive sense: by pre-specifying the nature of cooperative behavior, the metric-matching imitator has a much easier optimization problem, and more quickly demonstrates clear reciprocity to the innovator. The outcome is greater pro-sociality by all, and in a faster training time. To save compute, we terminated the metric-matching runs once they were consistently outperforming the niceness network. 

A representative episode from the trained agents in the niceness-network case shows the innovator taking a sustainable path through the apples, with imitators striving to match this.\footnote{A video is available at \url{https://youtu.be/pvdMt_0RCpw}.} Interestingly, the society comes relatively close to causing the tragedy of the commons. Intuitively, when apples are scarce, the actions of each agent should have a more significant effect on their co-players, leading to a better learning signal for reciprocity.

\subsection{Analysis}
In this section we examine the policies learned by the models, and the learning dynamics of the system.

\subsubsection{Influence between agents}

If our imitators are learning reciprocity, the the policy of the innovator should meaningfully influence the behavior of the imitators at the end of training. We demonstrate this for the Cleanup environment, by learning a GraphNet model that indirectly measures time-varying influence between entities \cite{tacchetti2018relational}. In Cleanup, the entities of interest are agents, waste and apples. 

The input to our model is a complete directed graph with entities as nodes. The node features are the positions of each entity and additional metadata. For waste and apples, the metadata indicates whether the entity has spawned. For agents, it contains orientation, last action and last reward, and a one-hot encoding of agent type (innovator or imitator). In addition, the graph contains the timestep as a global feature.

The model is trained to predict agent actions from recorded gameplay trajectories. The architecture is as follows: the input is processed by a GraphNet encoder block with $64$-unit MLPs for its edge, node and global nets followed by independent GRU units for the edge, node and global attributes with hidden states of size $32$, and finally a decoder MLP network for the agent nodes with layer sizes $(32, 32, 12)$. Importantly, the graph output of the GRU is identical in structure to that of the input. 

In \cite{tacchetti2018relational}, it was shown that the GRU output graph contains information about relationships between different entities. More precisely, the norm of the state vector along the edge from entity $A$ to entity $B$ computes the effective influence of $A$ on $B$, in the sense of Granger causality \cite{10.2307/1912791}. We may use this metric to evaluate the degree of reciprocity displayed by imitators.

Table \ref{tab:edges} shows the average norms of the state vectors along edges between imitators and innovators for our different imitation models, alongside an A3C baseline. The edge norm is greatest from the innovator to the imitator, strongly exceeding the baseline, indicating the innovator has a significant effect on the imitator's actions. The effect is strikingly visible when the output graph net edges are superimposed on a representative episode with metric-matching imitators, with thicknesses proportional to their norms.\footnote{A video is available at \url{https://youtu.be/NoDbUMkBfP4}. In the video, the innovator is purple, and the imitators are sky blue, lime, rose and magenta.}

\begin{table}
\begin{tabular}{|c|c|c|c|c|}
\hline
Experiment & In-Im & Im-In & Im-Im  & In-In \\ \hline
A3C & --- & --- &  --- & 0.27 \\ \hline
Metric matching, 4 A3C & \textbf{0.97} & 0.30 & 0.28 &  --- \\ \hline
Niceness network, 4 A3C & \textbf{0.35} & 0.21 & 0.22 &  --- \\ \hline
\end{tabular}
\caption{Edge norms for graph net models trained to predict future states and actions. For metric-matching and niceness network imitators, we see that the influence of the innovators on imitators is greater than for any other pair of agent types.}
\label{tab:edges}
\end{table}

\subsubsection{Ablation}
\begin{figure}
    \includegraphics[scale=0.4]{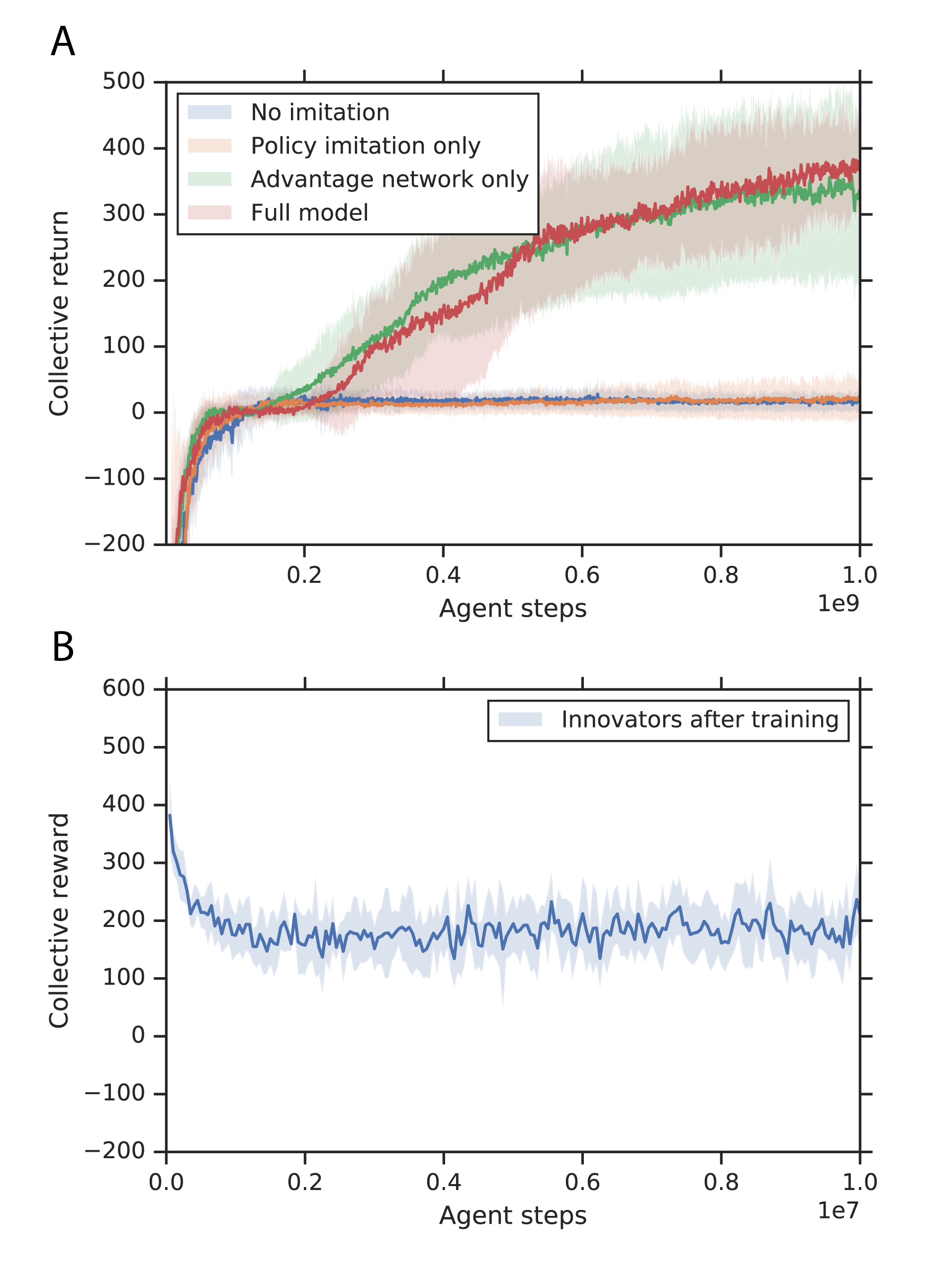}
    \centering
    \caption{(A) Ablating the model reduces performance. Policy imitation alone cannot resolve the social dilemma. Imitation reward alone resolves the dilemma inconsistently, since policy divergence may occur, destabilizing the niceness network. (B) Imitation is necessary to maintain cooperation. When innovators that have learned to be pro-social are co-trained, the group outcomes quickly degrade.}
    \label{fig:ablation}
\end{figure}

In the Cleanup environment, we examine the performance of the model with various components of the imitator ablated. We observe that with only the policy deviation cost, the performance is no better than with purely selfish agents. With the niceness network intrinsic reward, but no policy deviation cost, we see less consistent results across random seeds for the environment and agent (Figure \ref{fig:ablation}A).

\subsubsection{Instability of solution without imitation}
In the Cleanup environment, we take an innovator trained alongside four imitators, and run several copies of it in the environment, continuing to learn. We see that the contributions and collective reward quickly fall, as the innovators learn to be more selfish (Figure \ref{fig:ablation}B). This shows that for this environment, reciprocity is necessary not only to find cooperative solutions but also to sustain cooperative behaviour.

\subsubsection{Predictions of niceness network}
We analysed the predictions of the niceness network in the Coins environment, to determine whether the imitator has correctly captured which actions of the innovator are beneficial and harmful. We rolled out $100$ episodes using the final policies of the innovator and imitator from a run with the niceness network. On average, we found that the niceness network on average makes significantly negative predictions ($-0.35 \pm 0.02$) for actions which pick up the wrong coin, and near zero predictions for both picking up one's own coin ($-0.004\pm 0.004$) and actions which do not pick up a coin ($0.007 \pm 0.001$).

\begin{figure}
    \includegraphics[scale=0.5]{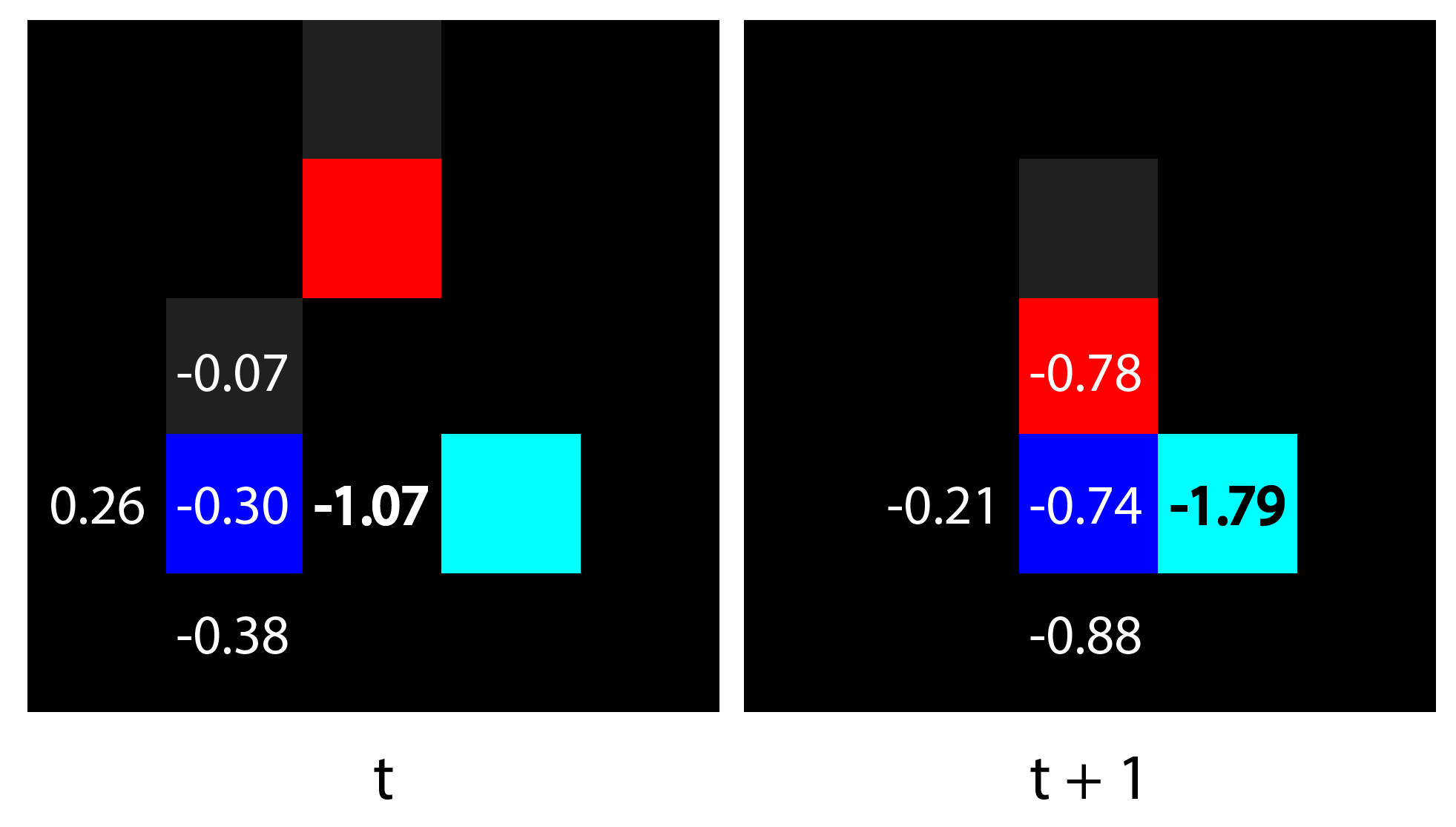}
    \centering
    \caption{The niceness network accurately measures the pro-sociality of different actions taken by the innovator. In this example, we see two consecutive frames from a test episode. The innovator (dark blue) starts two steps away from the imitator's coin (light blue). The $Q$-value for each action is indicated on each frame. It is lowest for actions that move the innovator closer to the coin (bold), and highest for actions that move the innovator away from the coin.}
    \label{fig:coin_advantages}
\end{figure}

On a more qualitative level, we display some of the predictions of the niceness network for the first of these episodes in figure \ref{fig:coin_advantages}. There we see that the niceness network predicts negative values for picking up the other agent's coins, and for actions which take the agent nearer to such coins.

\section{Discussion}

Our reciprocating agents demonstrate an ability to elicit cooperation in otherwise selfish individuals, both in $2$-player and $5$-player social dilemmas. Reciprocity, in the form of reciprocal altruism \cite{trivers}, is not just a feature of human interactions. For example, it is also thought to underlie the complex social lives of teleost fish \cite{Brandl2015}. As such, it may be of fundamental importance in the construction of next generation social learning agents that display simple collective intelligence \cite{DBLP:journals/corr/cs-LG-9908014}. Moreover, combining online learned reciprocity with the host of other inductive biases in the multi-agent reinforcement learning literature \cite{RePEc:zur:iewwpx:004, 2018arXiv181008647J, 2018arXiv180610071L, DBLP:journals/corr/abs-1710-06975, DBLP:journals/corr/LoweWTHAM17} may well be important for producing powerful social agents.

In the $5$-player case, our experiments place the innovator in the leadership role, giving rise to pro-sociality. An obvious extension would involve training several imitators in parallel, leading to a situation more akin to conformity; see for example \cite{doi:10.1146/annurev.psych.55.090902.142015}. In this case, all individuals change their responses to match those of others. A conformity variant may well be a better model of human behavior in public goods games \cite{BARDSLEY2005664}, and hence may generalize better to human co-players.

It is instructive to compare our model to the current state-of-the-art planning-based approach, approximate Markov Tit-for-Tat (amTFT) \cite{DBLP:journals/corr/LererP17}. There, reciprocity is achieved by first learning cooperative and defecting policies, by training agents to optimize collective and individual return respectively. The reciprocating strategy uses rollouts based on the cooperative policy to classify actions as cooperative or defecting, and responds accordingly by switching strategy in a hard-coded manner.

In the Coins environment, amTFT performs significantly better than both our niceness network and metric-matching imitators, solving the dilemma perfectly. We believe is because it better fulfills Axelrod's clarity condition for reciprocity. By switching between two well-defined strategies, it produces very clear responses to defection, which provides a better reinforcement learning signal driving innovators towards pro-sociality.

On the other hand, our model is more easily scalable to complex environments. We identify three properties of an environment which make it difficult for amTFT, but which do not stop our model from learning to reciprocate.
\begin{enumerate}
    \item If no perfect model of the environment exists, or rolling out such a model is infeasible, one must evaluate the cooperativeness of others online, using a learned model.
    \item The cooperative strategy for amTFT is learned on collective return. For games with multiple agents, this may not yield a unique policy. For example, in the Cleanup environment, the set of policies maximizing collective return involve some agents cleaning the river, while other eat apples.
    \item If cooperation and defection are nuanced, rather than binary choices, then to reciprocate you may need to adjust your level of cooperativeness to match that of your opponent. This is hard to achieve by switching between a discrete set of strategies.
\end{enumerate}

This leaves open an important question: how do we produce reciprocity which is both clear and scalable to complex tasks? One approach would be combining a model like ours, which learns what to reciprocate, with a method which switched between policies in a discrete fashion, as in the previous planning-based approaches of \cite{DBLP:journals/corr/LererP17, kleiman, DBLP:journals/corr/abs-1710-06975}. This might lead reciprocity algorithms which can be learned in complex environments, but which are clearer and so can induce cooperation in co-players even more strongly than our model.

%% file: appendix.tex
\appendix
\section{Hyperparameters} \label{appendix:arch}
In all experiments, for both imitators and innovators, the network consists of a single convolutional layer with $6$ output channels, a $3 \times 3$ kernel and stride $1$, followed by an two-layer MLP with hidden sizes $32$, an LSTM \cite{Hochreiter:1997:LSM:1246443.1246450} with hidden size $32$ and linear output layers for the policy and baseline. The discount factor for the return is set to $0.99$, and the learning rate and entropy cost were tuned separately for each model-environment pair.

The architecture for the niceness network is a non-recurrent neural network with the same convnet and MLP structure as the reinforcement learning architecture (though the weights are not shared between the two). The output layer of the MLP is mapped linearly to give outputs for $V^{\pi_{\textrm{inv}}}(s)$, $Q^{\pi_{\textrm{inv}}}(s, a)$ for each possible action, and $\hat\pi_{\textrm{inv}}(s, a)$ for each possible action.

For the niceness network, we used a separate optimizer from the RL model, with a separate learning rate. Both optimizers used the RMSProp algorithm \cite{rmsprop}. The hyperparameters used in each experiment are shown in Table \ref{table:hyperparameters}.
\begin{table*}
\begin{tabular}{|l|l|l|l|l|l|l|}
\hline
Hyperparameter & Coins (MM) & Coins (AN) & Cleanup (MM) & Cleanup (AN) & Harvest (MM) & Harvest (AN) \\ \hline
$c_{im}$ - imitation reward weight & $0.1$ & $0.02$ & $0.1$ & $0.1$ & $3.0$ & $0.1$ \\ \hline
$\gamma$ - imitation memory decay & $0.95$ & $0.95$ & $0.95$ & $0.95$ & $0.95$ & $0.95$ \\ \hline
$c_{KL}$ - policy imitation weight & -- & $0.03$ & -- & $0.1$& -- & $0.1$ \\ \hline
Entropy weight & $0.003$ & $0.003$ & $0.001$ & $0.001$ & $0.003$ & $0.003$ \\ \hline
RL learning rate & $0.0005$ & $0.0001$ & $0.0001$ & $0.0001$ & $0.0005$ & $0.0001$ \\ \hline
Advantage network learning rate & -- & $0.001$ & -- & $0.0001$ & -- & $0.0001$ \\ \hline
Advantage network TD-$\lambda$ & -- & $0.9$ &  -- & $0.95$ &  -- & $0.95$ \\ \hline
Advantage network discount factor & -- & $0.9$ & -- & $0.99$ & -- & $0.99$ \\ \hline
\end{tabular}
\caption{Hyperparameters used in metric matching and advantage network imitation experiments.}
\label{table:hyperparameters}
\end{table*}